\documentclass{revtex4}
\usepackage{amsmath}
\usepackage{graphicx}
\newcommand{\be}{\begin{equation}}
\newcommand{\ee}{\end{equation}}
\begin{document}
\title{Observations Outside the Light-Cone: Algorithms for Non-Equilibrium
and Thermal States}
\author{M.~B.~Hastings$^{1,2}$}
\affiliation{$^1$ Center for Nonlinear Studies and Theoretical Division,
Los Alamos National Laboratory, Los Alamos, NM, 87545 \\
$^2$ Kavli Institute of Theoretical Physics, University of California,
Santa Barbara, CA, 93106}
\begin{abstract}
We apply algorithms based on Lieb-Robinson
bounds to simulate time-dependent and thermal quantities in quantum
systems.  For time-dependent systems, we modify
a previous mapping to quantum circuits to significantly reduce the computer
resources required.  This modification
is based on a principle of ``observing" the system outside the light-cone.
We apply this method to study spin relaxation in systems
started out of equilibrium with initial conditions that give rise to
very rapid entanglement growth.  We also show that it is possible to
approximate time evolution under a local Hamiltonian by
a quantum circuit whose light-cone naturally matches the Lieb-Robinson
velocity.
Asymptotically, these modified methods allow a doubling of the system size that
one can obtain compared to direct simulation.
We then consider a different problem of
thermal properties of disordered spin chains and use quantum belief propagation
to average over different configurations.  We test this
algorithm on one dimensional systems with mixed ferromagnetic and
anti-ferromagnetic bonds, where we can compare to quantum Monte Carlo, and
then we apply it to the study of disordered, frustrated spin systems.
\end{abstract}
\maketitle
\section{Introduction}
Matrix product and density-matrix renormalization group methods
provide one of the most powerful ways of simulating one-dimensional
quantum systems.  In addition to ground state properties\cite{mpsg}, they
have been extended to thermal and open systems\cite{mpst} and dynamical
problems\cite{tebd}.  The reason for the success of these algorithms is
that in many cases the appropriate quantum state can be well-approximated by
a matrix product state, giving a very compact representation of the
state of the system.  

In some cases, we even have theorems that quantify the
accuracy of matrix product states.
For ground states of local quantum systems with
a gap, the ability to represent ground states as matrix product states follows
from bounds on the entanglement entropies\cite{areal}.  Related results
are available for thermal systems\cite{solvgap,thermalarea} and 
for non-equilibrium
states obtained by starting with a factorized state and evolving under
a local Hamiltonian for a time $t$; in the first case
the bond dimension needed
to get a good matrix product approximation to the desired state
 scales exponentially in $\beta$, while in the second case it
scales exponentially
in $t$\cite{qcirc}.  The two results \cite{solvgap,qcirc} are constructive
proofs, which give an algorithm to find the matrix product state.
All of these constructive proofs rely heavily on Lieb-Robinson bounds\cite{lr1,lr2,lr3,lr4}.

In contrast to these constructive proofs,
the matrix product algorithms used in practice are variational: they
involve optimizing over different matrix product states to find the
best one.  This works well because in many practical cases the
entanglement grows much more slowly than the upper bounds set by theory.
For example, systems described by conformal field theory have an entanglement
entropy growing only logarithmically with system size\cite{cftent}, while
the area law bound\cite{areal} gives no useful result for these systems
due to the absence of a gap.  Similarly, for many initial conditions,
evolution under a local Hamiltonian gives an entanglement entropy growing
only logarithmically in time\cite{logtime,logtime2,schl} while the theoretical
upper bound gives an entanglement entropy growing linearly in time\cite{lrwall}.

In this paper, we argue that there are many situations in which
algorithms based on Lieb-Robinson bounds are the best
technique.
We first look at the case of time evolution in systems out of equilibrium.
Here, there are initial conditions for which the entanglement entropy
is known to grow linearly in time\cite{calab}, in accordance with conformal
field theory predictions\cite{cftrapid}.  Roughly speaking, logarithmic
entropy growth tends to occur in cases where we can divide the chain
into a small number of subchains such that the initial state is an eigenstate
of the Hamiltonian on each subchain; for example, starting an XXZ spin chain
in a state in which all the spins on the left half of the chain are up and
all the spins on the right half of the chain are down leads to a logarithmic
entropy growth\cite{schl}.
On the other hand,
the linear entropy growth tends to occur in cases where the initial state
differs from an eigenstate of the Hamiltonian on every subsystem of the full
spin chain.  For example, starting an XXZ spin chain in an initial condition 
in which the spins alternate between up and down (a Neel state, the ground
state when the Ising term is the only term in the Hamiltonian) leads to a
linear entropy growth\cite{calab}.

For system with linear entropy growth,
matrix product methods will require a bond dimension
growing exponentially in time to obtain accurate results.  At this point,
both variational matrix product and
constructive, Lieb-Robinson-based methods require
resources growing exponentially in time.  The question, then, is how to obtain
the smallest exponential.  To some extent, the Lieb-Robinson based methods
(such as \cite{qcirc} and the methods below) are ``worst-case": the
bond dimension depends on theoretical upper bounds for {\it arbitrary} local
Hamiltonians, while the matrix product methods can adaptively find better
representations.  On the other hand, there are some disadvantages to
matrix product methods.  To get a rough idea of the resources required,
let us consider a system of $N$ spins, each of spin-$1/2$, with a local
Hamiltonian.  The simplest algorithm to simulate this system for a time
$t$ involves writing
the initial condition down in some basis and then directly simulating it
(we discuss below different methods for doing this), requiring resources
scaling as $2^N t$ using sparse matrix methods.
A matrix product algorithm can avoid
truncation error for this system by using matrices with
bond dimension $2^{N/2}$\cite{2n2,2n2why}.
However, the algorithm must then perform
singular value decompositions and eigenvalue calculations, taking a time which
scales as the cube of these matrices, and hence of order $2^{3N/2}$, which
is slower.  Of course, the matrix product methods are really only useful
if they are able to represent the system with a smaller bond dimension.
In this case, though, even if such a representation exists, the algorithm
must find it, and this can pose a problem.
The algorithm for simulation of non-equilibrium systems depends on
breaking the time evolution into a series of small Trotter steps; if the
Trotter steps are too long this can lead to Trotter error, while if they
are too short, the truncation error can grow rapidly: even if there is
a good state the algorithm may not find it\cite{schl}.

We can use this scaling of the difficulty with $N$ to get an
idea of the scaling of computation effort with time, using
a Lieb-Robinson
bound on the group velocity, $v_{LR}$.  Suppose we wish to compute
the expectation value of a local observable, such
as a spin on a site, at a time $t_f$, starting from a factorized
state at time $t=0$.  The number of spins
in the past light-cone of this spin is $2 v_{LR} t$, and the Lieb-Robinson
bounds imply that the effect of spins outside the light-cone is exponetially
small.  
Thus, it suffices to simulate only the dynamics of the
$N=2 v_{LR} t_f+O(\log(\epsilon))$
spins closest to the given spin in order to compute the expectation
value to an accuracy $\epsilon$.  This requires an effort scaling exponentially
in time as $t^2 2^{2 v_{LR} t}$.
Similarly, if the entanglement entropy grows linearly in time,
the matrix product methods also require an effort scaling exponentially
in time.

Our main result in this paper is the light-cone quantum circuit algorithm,
an application of the Lieb-Robinson bound that
allows us to simulate the evolution of local observables 
with resources growing 
asymptotically as only $N t 2^{v_{LR} }$, by some statistical sampling.  
This allows twice as large systems as the direct method.  We analyze
the entropy growth in these systems and argue that matrix product methods
are also less efficient for long time simulation.  We apply
the light-cone quantum circuit algorithm
then to the problem of spin relaxation in spin chains started
in the Neel state.
The physical idea behind the light-cone quantum
circuit algorithm is as follows: to find the state of
a given spin at a time $t_f$, we only
have to track the dynamics within the past light-cone of the spin.  For times
$t$ close to zero, the past light-cone includes roughly $2 v_{LR} t_f$ spins,
but at these early times the entanglement is small and hence the
computational effort should be less.  For times $t$ close to $t_f$,
the past light-cone includes few spins and hence should be easier to simulate.

The paper is organized as follows.  We first derive the light-cone
quantum circuit algorithm.  We then apply it to spin
relaxation, and study oscillations of the central spin, decay of the envelope
of the oscillations, and also seemingly random oscillations of the central
spin in chains where boundary effects become important.  We then derive
a related quantum circuit method, the corner transfer quantum circuit which
may be useful for studying the evolution of global observables in 
highly entangled non-equilibrium states.

We then turn to a different problem, 
presenting one other application of Lieb-Robinson methods,
using the quantum belief propagation algorithm\cite{qbp} to study thermal
states in disordered systems.  The quantum belief propagation algorithm
explicitly constructs a matrix product state for a thermal quantum system.
While it manipulates operators, rather than states, and thus can be
computationally expensive, it has other advantages.  It has
no Trotter error, making
it fast and accurate at high temperatures: it can obtain quantities such as
the susceptibility peak to higher accuracy using fewer resources than
methods such as transfer matrix DMRG\cite{tdmrg}, although at low temperatures
it breaks down, with the resources required scaling exponentially with the
temperature.  In this case, the exponential scaling with the temperature
is again related to a linear relationship between a time scale, in this
case $\beta=1/T$, and a length scale.
It can be applied to random systems, where transfer matrix DMRG cannot be
used because of a lack of translation invariance.  A good test of
variational matrix product\cite{mpst}
methods on this kind of system is lacking, so we 
cannot compare here.  We apply the quantum belief propagation algorithm
to two random systems, one without frustration where we can compare to quantum
Monte Carlo and one with frustration where Monte Carlo methods are not
applicable.

\section{Quantum Circuit Methods}

In this section we present the various quantum circuit methods.  We begin
by reviewing previous work, and then derive the light-cone
quantum circuit algorithm, and apply it to a problem of spin relaxation.
We then use previous results on the entanglement entropy growth to estimate the
computational resources required for different approaches to this problem,
and finally we present the corner transfer quantum circuit method,
an extension which allows access to global quantities.

\subsection{Background}
To understand our algorithm, we first review the ideas in \cite{qcirc}
which gives a construction of a matrix product operator approximation
to the time evolution operator, $\exp(-i H t)$, using resources
exponential in $t$.
We consider a local Hamiltonian
\be
H=\sum_i h_i,
\ee
where each $i$ acts on
sites $i,i+1$.  This Hamiltonian obeys a Lieb-Robinson bound: given
any operator $O$ which has support on set of sites $X$, the operator
$\exp(i H t) O \exp(-i H t)$ can be written, with exponentially small
error, as an operator acting on the set of sites $i$ within distance
$v_{LR} t$ of $X$, where $v_{LR}$ is the Lieb-Robinson group velocity.
 
To simulate the system for a time $t$,
we divide the system into blocks of length $l$, where $l$ is
slightly larger than $2 v_{LR} t$ (the error in the approximation
will be exponentially small in $l-2v_{LR} t$).
We then let $H=H_0+H'$ where $H_0$ is the sum
of the Hamiltonians on each block and $H'$ is the Hamiltonian connecting
the blocks:
\be
H_0=\sum_{k}\sum_{i=kl+1}^{i\leq (k+1)l-1} h_i,
\ee
\be
H'=\sum_k h_{kl},
\ee
where the sum ranges over integers $k$.

We then write
\be
\exp(-i H t)=\Bigl( {\cal T} \exp[-i \int_0^t \exp(-i H_0 t') H' \exp(i H_0 t')]
\Bigr)
\exp(-i H_0 t),
\ee
where ${\cal T}$ denotes that the exponential is time-ordered.
The operator $\exp(-i H_0 t)$ is equal to the product $\prod_k U_k$, where
$U_k$ is a unitary operator acting on sites $kl+1,kl+2,...,kl$:
\be
U_k=
\exp(-i \sum_{i=kl+1}^{i\leq (k+1)l-1} h_i t).
\ee
Using the Lieb-Robinson bounds, we can approximate
$\exp(-i H_0 t') h_{kl} \exp(i H_0 t')$ by an operator $h_{kl}(t')^{loc}$
which has support on
sites $kl-l/2+1,...,kl+l/2-1$ for $|t'|\leq t$ and for $l>v_{LR} t$.
Then the operator
${\cal T}\exp[-i\int_0^t \exp(-i H_0 t') H' \exp(i H_0 t')]$ can
then be approximated by a product $\prod_k V_k$ where
\be
V_k=
{\cal T} \exp[-i\int_0^t h_{kl}(t')^{loc}].
\ee

\begin{figure}
\centerline{
\includegraphics[scale=0.7]{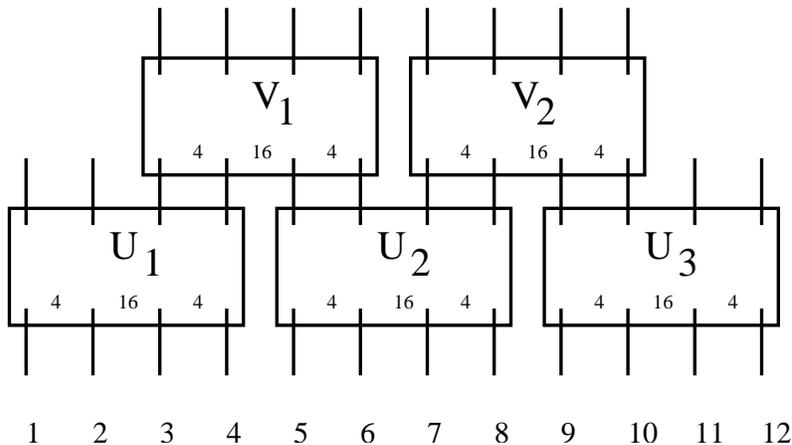}}
\caption{Support of the operators $U_k,V_k$ and the quantum circuit
for $N=12$, $l=4$.  The small numbers at the bottom of operators $U_k,V_k$
represent the bond dimensions required to represent these operators as
matrix product operators; the maximum product of these across any bond
is $16=4^{l/2}$.}
\label{figqc}
\vspace{5mm}
\end{figure}

The key in this construction is that the intervals
$kl-l/2+1,...,kl+l/2-1$ do not overlap for different $k$.
This construction expresses the time evolution as
a quantum circuit:
\be
\label{approx}
\exp(-i H t)\approx\prod_k V_k \prod_k U_k.
\ee
The support of the operators $U_k,V_k$ and the circuit is shown in
Fig.~\ref{figqc}.

Precise error bounds can be given using Lieb-Robinson bounds for
the error in Eq.~(\ref{approx}).  To get an error of order $\epsilon$ in
the propagator (\ref{approx}), we only need to take $l=
2v_{LR}t+{\cal O}(\log(N/\epsilon))$.  In what follows, we will not make
detailed error estimates, since Lieb-Robinson error estimates are fairly
simple and are by now standard in the literature; when we say that it suffices
to take a length scale ``of order" $v_{LR} t$ to obtain an approximation
to a given local quantity, we mean that by taking the length scale
$v_{LR}t+O(\log(N/\epsilon))$ we can obtain an error of order $\epsilon$
in the state; when we are computing expectation values of local quantities,
to obtain an error of order $\epsilon$ we need a length scale
$v_{LR}t+O(\log(v_{LR}t/\epsilon))$, so that the error bound does not
depend on $N$ in this case.

Suppose we want to apply the quantum circuit procedure
to compute the time evolution of some state $\Psi_0$.  For
simplicity, let $\Psi_0$ be a factorized state (later we
discuss the case where $\Psi_0$ is a
matrix product state both in this procedure and
using our algorithm, and we find that using the idea of ``observation" discussed
below the case of matrix product state initial
conditions presents no additional difficulty).
The operator $U_k$ is an operator acting on $2l$ sites.  Any such operator
can be written as a matrix product operator with a bond dimension equal to
$4^{2l/2}=2^{2l}$\cite{2n2}.
This maximum bond dimension is achieved halfway across the interval of
length $2l$, while at any point a distance $d$ from one of the ends
of the interval the bond dimension is only $2^{d}$, as shown in
Fig.~\ref{figqc}.

The same holds for $V_k$, and so the maximum product of bond dimensions across
any bond is $2^{2l}$ (a slightly worse estimate of $2^{4l}$ was found for
this construction in \cite{qcirc} since the fact that the bond dimension
may vary with position was not taken into account).

There are several problems, however, with implementing the above method
in practice, and it is these problems which we overcome with our
methods, the light-cone quantum circuit method and the
corner transfer quantum circuit method.  The first method is most appropriate
for computing local quantities (such as a spin or energy expectation value)
while the second method is most appropriate for finding a good global
approximation to the ground state.

The first problem is that operator equations of motion are computationally
expensive in practice.  To this end, we will modify the procedure to
deal only with state vectors, rather than operators.  The second problem
is that the velocity of the quantum circuit does not obviously match
the Lieb-Robinson velocity, in the following sense: given {\it arbitrary}
operators $U_k,V_k$, supported as described above, the product
$\prod_k V_k \prod_k U_k$ can propagate information by a distance of
$2l$ in each time step.  Since $l$ is roughly $v_{LR}t$, this means that
such a quantum circuit could have an effective velocity roughly
$2v_{LR}$.  Of course, the operators $U_k,V_k$ are not arbitrary operators,
but still we would like to fix this problem; we will show how to
do this with the corner transfer quantum circuit method below which also
leads to improved estimates on the maximum matrix product state dimension
needed.

However, the real problem with this method is that it doesn't lead to
any improvement over naive simulation when it comes to computing local
observables.  The main problem we consider in this section is the
following: we start a spin chain at time $t=0$ in a factorized state $\Psi_0$,
and then evolve under a local Hamiltonian for to a final time
$t_f$, at which point
we wish to compute some local observable, such as $S^z_i$, the $z$-expectation
value of spin $i$.  
By the Lieb-Robinson bounds, we can approximate $S^z_i(t)$ by
considering only a subchain of the full chain: we consider only
sites $i-l,...,i+l$, where $l$ is slightly larger than $v_{LR} t_f$.
We then define $\Psi'$ to be the appropriate factorized state
on this subchain and evolve $\Psi'$ for a time $t_f$ using the
Hamiltonian $H'$ acting on the subchain.  We then compute
$\langle \Psi' | \exp(i H' t_f) S^z_i \exp(-i H' t_f)|\Psi'\rangle$.
Using sparse matrix methods to
compute the time evolution of $\Psi'$, this requires a computational
effort of order $l 2^{2l}$, which scales as $2^{2 v_{LR} t_f}$.  The
quantum circuit method discussed above would also require simulations
on intervals of length $2l$, and hence leads to no improvement
when computing this local quantity.

\subsection{Light-Cone Quantum Circuit Algorithm}
We now show how to
reduce the computational effort to an amount of order $2^{v_{LR} t_f}$,
allowing the time scale to be twice as large, by combining Lieb-Robinson
bounds with statistical sampling.  While we focus on this section on
starting in a factorized state, we later discuss the case of starting
in a matrix product state.

\begin{figure}
\centerline{
\includegraphics[scale=0.7]{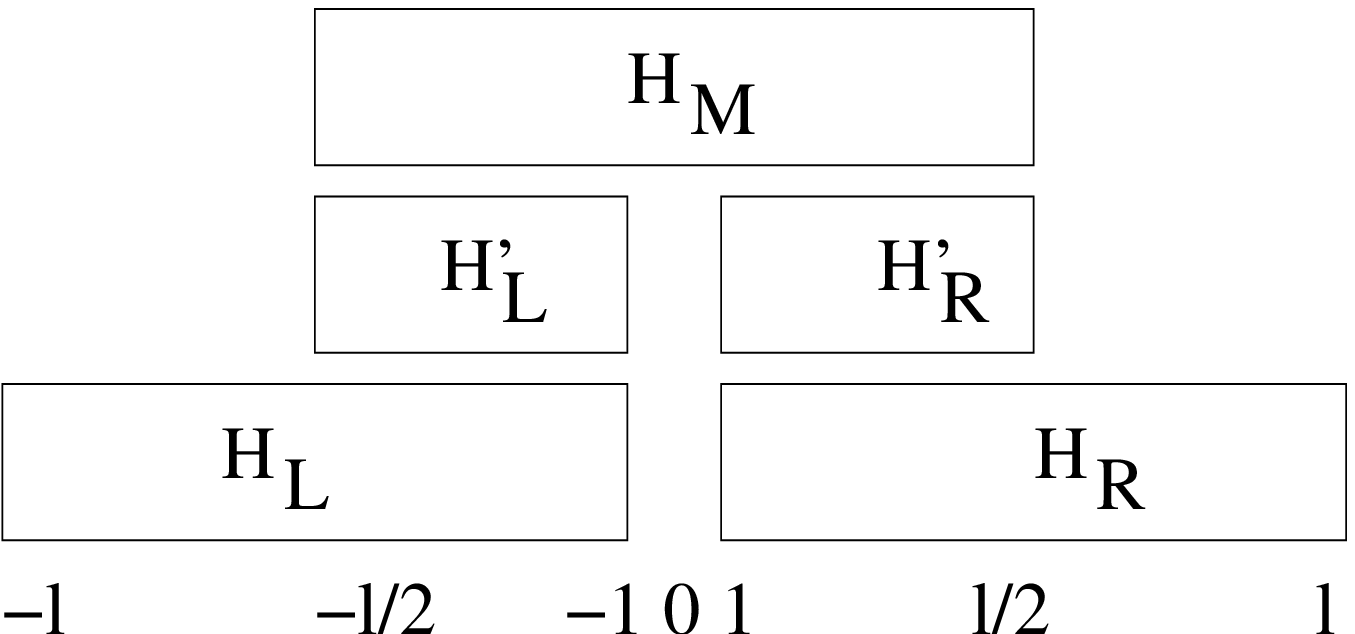}}
\caption{Support of the operators $H_L,H_R,H_M,H_L',H_R'$.}
\label{fighs}
\vspace{5mm}
\end{figure}

We now derive our algorithm, which we call the light-cone quantum circuit
method as it avoids keeping track of certain degrees of freedom outside
the light-cone by making certain
``observations" to reduce the computational
effort.
We first define a subchain of
length $2l+1$ and an initial state $\Psi'$ on that subchain
as above.  We label the sites in
the subchain by $-l,...,0,...,+l$.  We let $H'$ be the Hamiltonian
on the subchain and we write
\be
H'=H_L+H_R+H_B,
\ee
where $H_L$ acts on the left half of the chain (sites $-l,...,-1$),
$H_R$ acts on the right half of the chain, and
the boundary Hamiltonian $H_B$ acts on sites $-1,0,1$.  Thus, $[H_L,H_R]=0$.
We define $H_M$ to act on the {\it middle} half of the chain: it is supported
on sites $-l/2,...,+l/2$ (we pick $l$ even for simplicity).  See
Fig.~\ref{fighs}.

Using the Lieb-Robinson bounds in the same way
as in the quantum circuit method
above we can approximate the time evolution for a time 
$t_i=t_f/2$ by:
\be
\label{decom}
\Psi'(t_f/2)=
\exp(-i H' t_f/2)\Psi'\approx \exp(-i H_M t_f/2) \exp[-i (H_M-H_B) t_f/2]
\exp[i H_L t_f/2] \exp[i H_R t_f/2] \Psi'.
\ee
Note that the decomposition (\ref{decom}) is only good for times
of order  $t_f/2$.  We wish to compute the expectation value of $S^z_0$,
or any other observable on site $0$, at time $t_f$.
Again using the Lieb-Robinson bounds, the expectation value of this
is approximately equal to
\be
\label{middecom}
\langle\Psi'(t_f/2)|\exp(i H_Mt_f/2) O \exp(-i H_M t_f/2)|\Psi'(t_f/2)\rangle.
\ee
Combining Eqs.~(\ref{decom},\ref{middecom}) and our use of the Lieb-Robinson
bounds to approximate the expectation value of $O$ on the full chain by
its expectation value on the subchain, we have
\be
\langle \Psi_0|O(t)|\Psi_0\rangle\approx
\langle \tilde\Psi|O|\tilde \Psi\rangle,
\ee
where
\be
\tilde\Psi=\exp(-i H_M t_f/2) \Psi' \approx
\exp(-i H_M t_f) \exp[-i (H_M-H_B) t_f/2]
\exp[i H_L t_f/2] \exp[i H_R t_f/2] \Psi'.
\ee
The operator $H_M-H_B$ is a sum of two operators, $H_L'$ and $H_R'$,
where $H_L'$ acts on sites $-l/2,...,-1$ and $H_R'$ acts on sites
$+1,...,+l/2$.  Therefore,
\be
\label{tildPsi}
\tilde\Psi=\exp(-i H_M t_f/2) \Psi'\approx \exp(-i H_M t_f)
\Bigl( \exp[i H_L' t_f/2]
\exp[-i H_L t_f/2] \Bigr)
\Bigl( \exp[i H_R' t_f/2]
\exp[-i H_R t_f/2] \Bigr)
\Psi'.
\ee

Eq.~(\ref{tildPsi}) is not yet useful computationally, since it will require
an effort of order $l t_f 2^{2l}$ to compute the evolution of the state $\Psi'$.
We now describe the light-cone quantum circuit method to compute the
expectation value:
first, write
\be
\Psi'=\Psi_L\otimes \Psi_C \otimes \Psi_R,
\ee
where $\Psi_L,\Psi_R$ are states on the left and right half of the chain,
and $\Psi_C$ is a state on the center site.  
We compute the states 
\be
\label{psiLp}
\Psi_L'=(\exp[i H_L' t_f/2] \exp[-i H_L t_f/2] \Bigr)\Psi_L,
\ee
and
\be
\label{psiRp}
\Psi_R'=(\exp[i H_R' t_f/2] \exp[-i H_R t_f/2] \Bigr)\Psi_R,
\ee
which requires an effort of order $l t_f 2^{l}$.
Next, we introduce a complete orthonormal basis of states
on the sites $-l,...,-l/2-1$, which we label $\phi_L(\alpha)$, 
and another
a complete basis of states on sites $l/2+1,...,+l$ labelled
$\phi_R(\alpha)$.  
Then we decompose $\Psi_L'$ and $\Psi_R'$ as:
\begin{eqnarray}
\label{Alpha}
\Psi_L'=\sum_{\alpha} A(\alpha) \phi_L(\alpha)\otimes\xi_L(\alpha),
\\ \nonumber
\Psi_R'=\sum_{\alpha} A(\alpha) \phi_R(\alpha)\otimes\xi_R(\alpha),
\end{eqnarray}
where $\xi_L(\alpha)$ is some normalized state on sites $-l/2,...,-1$ and
$\xi_R(\alpha)$ is some normalized state on sites $+1,...,+l/2$.
The states $\phi_L(\alpha)$ need not be eigenvectors of any reduced
density matrix and the
states $\xi_L(\alpha)$ need not be orthogonal to each other, as
Eq.~(\ref{Alpha}) is {\it not} a Schmidt decomposition.
Thus, from Eqs.~(\ref{tildPsi},\ref{psiLp},\ref{psiRp},\ref{Alpha}),
\be
\label{statesample}
\langle \tilde\Psi|O|\tilde \Psi\rangle=
\sum_{\alpha_L} \sum_{\alpha_R} |A(\alpha_L)|^2 |A(\alpha_R)|^2
E(O,\alpha_L,\alpha_R),
\ee
where
\be
\label{heart}
E(O,\alpha_L,\alpha_R)=
\langle \xi_L(\alpha_L)\otimes \Psi_C \otimes \xi_R(\alpha_R)|
\exp(i H_M t_f) O \exp(-i H_M t_f)
|\xi_L(\alpha_L)\otimes \Psi_C \otimes \xi_R(\alpha_R)\rangle.
\ee
Eq.~(\ref{statesample}) is at the heart of the light-cone quantum circuit
approach.  Numerically we proceed as follows: first, we compute
$|A(\alpha_L)|^2$ and $|A(\alpha_R)|^2$ for all $\alpha_L,\alpha_R$.  This
requires an effort of order $2^{l}$.
We then do a statistical sampling: we randomly pick an $\alpha_L$ and
an $\alpha_R$ according to the probability distributions
$|A(\alpha_L)|^2$ and $|A(\alpha_R)|^2$, and compute the average
$E(O,\alpha_L,\alpha_R)$.  We repeat this procedure many times,
to average over different choices of $\alpha_L,\alpha_R$.

The computational effort required then scales as only $l 2^{l}$, or
roughly $t 2^{v_{LR} t}$ as claimed.  Asymptotically this allows double the
time.  The time scales linearly in the number of iterations of statistical
sampling, which we denote $N_{it}$.  However, if $O$ has bounded operator
norm, then $E(O,\alpha_L,\alpha_R)$ has bounded moments, and
so by the central limit theorem, the number of iterations required still scales
only polynomially in the error.

\subsection{Results on Non-Equilibrium Dynamics}

We now discuss results from this method, as well as some implementation
details.
We consider evolution under the XXZ Hamiltonian
\be
\label{xxz}
H=\sum_i (S_i^x S_{i+1}^x + S_i^y S_{i+1}^y+\Delta S_i^z S_{i+1}^z).
\ee
For $\Delta=0$, this problem can be mapped to free fermions by
a Jordan-Wigner transformation and solved exactly.  We use this as
a check on our results later.

We used as a starting point the Neel state, with spins alternating
up and down, and we computed the time dependence of $S^z(t)$ for
the central spin.  The main numerical effort is to compute the 
evolution of a state under a Hamiltonianm, which we did using
a combination of short steps with
a series method.
For example, to compute
$\exp(-i H_L t)\Psi_L$
we divide the time $t$ into shorter intervals of time $t_0$, and
compute $\exp(-i H_L t_0)\Psi_L=\Psi_L-it_0 H_L \Psi_L
-(t_0^2 H_L^2/2!) \Psi_L+...$, keeping a
fixed number of terms in this series.  We then repeat this procedure $(t/t_0)$
times.  To obtain negligible error
for a chain of $20$ sites with $t_0=1$ required going
to roughly $40$-th order for $|\Delta|\leq 1$, while for $\Delta=2$ slightly
longer series were required.  A more sophisticated way of doing the time
evolution would be to build a tridiagonal Hamiltonian in the Krylov
space spanned by $\Psi_L,H_L \Psi_L,...,H_L^k\Psi_L$ for some $k$, and
then evolve exactly with this Hamiltonian\cite{spec}.

Another important point of numerical simulation is the use of symmetries.
We can choose the states $\phi_L(\alpha)$ and $\phi_R(\alpha)$ to be
eigenstates of total $S^z$.  Then, since the state $\Psi'$ is an eigenstate
of total $S^z$, the state
$\xi_L(\alpha_L)\otimes \Psi_C \otimes \xi_R(\alpha_R)$ is also an
eigenstate of total $S^z$, which allows us to use symmetries when computing
the evolution of the state.  Since most of the numerical time is consumed
statistically sampling
$E(O,\alpha_L,\alpha_R)$, we build the sparse matrix for the Hamiltonian
$H_M$ in each spin sector once, before doing the sampling, and then
run the sampling.

It is also possible, although we did not implement it, to take into
account reflection symmetry.  Since $\alpha_L$ and $\alpha_R$ are
chosen independently, the state
$\xi_L(\alpha_L)\otimes \Psi_C \otimes \xi_R(\alpha_R)$ does not have
reflection symmetry.  However, if both the Hamiltonian $H_M$ and the
operator $O=S^z_i$ have reflection symmetry, then it is useful to
write
\be
\xi_L(\alpha_L)\otimes \Psi_C \otimes \xi_R(\alpha_R)=
\Psi_S(\alpha_L,\alpha_R)+\Psi_A(\alpha_L,\alpha_R),
\ee
where $\Psi_S,\Psi_A$ are symmetric and anti-symmetric states.
Then,
\begin{eqnarray}
\label{statsamp}
E(O,\alpha_L,\alpha_R)=&&
\langle \Psi_S(\alpha_L,\alpha_R|
\exp(i H_M t) O \exp(-i H_M t)| \rangle \Psi_S(\alpha_L,\alpha_R)
\rangle\\ \nonumber &+&
\langle \Psi_A(\alpha_L,\alpha_R|
\exp(i H_M t) O \exp(-i H_M t)| \rangle \Psi_A(\alpha_L,\alpha_R)
\rangle,
\end{eqnarray}
and so we can statistically sample one of the two terms on the right-hand side
of Eq.~(\ref{statsamp}).  Note that on each iteration we randomly choose
an $\alpha_L$ and an $\alpha_R$ and then randomly choose a term in
Eq.~(\ref{statsamp}), rather than repeatedly sampling Eq.~(\ref{statsamp}).

As the algorithm is described above,
the initial computation of the states $\Psi_L(\alpha)$ and $\Psi_R(\alpha)$
depends on the final time.  For each final time $t$, we have to compute
a new set of states $\Psi_L(\alpha)$ and $\Psi_R(\alpha)$ and then do
the statistical sampling.  However, in fact, we can speed the algorithm
up at a slight cost in accuracy: we fix a given $t_f$ and on each
statistical sample we compute the
state 
\be
\exp(-i H_M t_f)
|\xi_L(\alpha_L)\otimes \Psi_C \otimes \xi_R(\alpha_R)\rangle,
\ee
to evaluate the expectation value in Eq.~(\ref{heart}).
We then act on this state with the operator $\exp(i H_M \delta t)$
for some small $\delta t$.  We then use this new state to compute
an approximation to the expectation value at time $t_f-\delta t$.
We then act on that state with $\exp(-i H_M \delta t)$ to compute
an approximation to the expectation value at time $t_f-2 \delta t$,
and so on.  Since the computational cost of performing the time
evolution of a state under a Hamiltonian is proportional to the time evolved,
these additional steps are relatively cheap, for $\delta t<<t_f$.
There is a small cost in accuracy: in general, to compute expectation
values at a time $t$, we can do the initial evolution for a time $t_i$, and
then evolve further for time $t-t_i$.  To make the effect of boundary
conditions as small as possible, we would like to have both $t_i$ and $t-t_i$
as small as possible, which is why above we chose to evolve for a time
$t_i=t_f/2$.  However, if $\delta t<<t_f$, then we are not far away
from the ideal choice of $t_i$ by initially evolving for time $t_f/2$ and
then evolving for time $t_f/2-\delta t$.
We followed this procedure in the numerical work below, with $\delta t=0.25$
and taking $t_f$ to be spaced with integer steps using $t_0=1$.
This accounts for
some of the slight kinks in the curves after every integer value of $t$.

The spin-wave velocity of Hamiltonian (\ref{xxz}) for $\Delta \leq 1$
is given by
\be
v_{sw}=(\pi/2) \sin(\theta)/\theta,
\ee
where $\cos(\theta)=\Delta$\cite{swvel}.
On the other hand,
in the development above, we used a Lieb-Robinson velocity $v_{LR}$, where
\be
\label{ineqV}
v_{LR}\geq v_{sw}.
\ee
Using the Lieb-Robinson bounds, we showed that we could accurately simulate
for a time $t$ using length scales $l=v_{LR} t$.  As mentioned above, we
do not give precise error estimates, but it is not hard to give
rigorous estimates of the error.
However, the Lieb-Robinson bound is actually fairly conservative
because of Eq.~(\ref{ineqV}), and thus in practice length scales
$v_{sw}t$ suffice to get good results.

\begin{figure}
\centerline{
\includegraphics[scale=0.5]{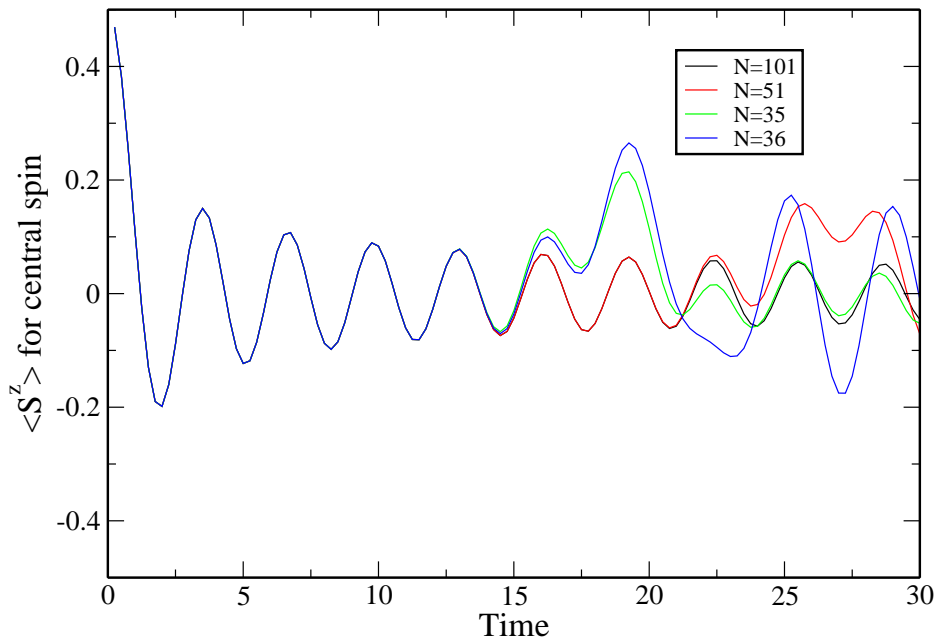}}
\caption{Time dependence of $\langle S^z \rangle$ for the central spin as
a function of time.
Curves are $N=101$ (black), $N=51$ (red), $N=35$ (green), and
$N=36$ (blue).
}
\label{XY}
\vspace{5mm}
\end{figure}

We begin by illustrating results for the XY chain, with $\Delta=0$.
In Fig.~\ref{XY}, we illustrate results from exact simulations
of the XY chain for various sizes.  We consider chains with open
boundary conditions with $N=35,51,101$ and a chain with periodic boundary
conditions with $N=36$.  For the chains with open boundary conditions,
we plot the average of the central spin as a function of time, while for
the periodic boundary conditions we plot the average of an arbitrarily
chosen spin as a function of time.  The large exact simulations are possible
because this chain can be mapped to free fermions by a Jordan-Wigner
transformation.  Later we will present comparison of these results
to light-cone quantum circuit results.  For now, we discuss aspects
of Fig.~\ref{XY} which show the influence of boundary effects.
In the range of times, the chain with $N=101$ shows no effect of the
boundary.  There are oscillations of the spin with frequency $\omega$
roughly $2$, with an envelope decaying as $1/\sqrt{t}$ so
as
\be
\label{envelope}
\langle S_i^z\rangle\sim \frac{1}{\sqrt{t}} \cos(\omega t+\theta_0),
\ee
for $\theta_0\approx (3/4) \pi$.
Simulations on longer chains show that as long as 
$N$ is less than $t/2$, the decaying oscillations of
Eq.~(\ref{envelope}) continue to hold.  In regard to the $1/t^{1/2}$
decay of oscillations found here numerically, it is interesting to
note a similar power-law decay found for a different system of free
bosons, where fluctuations about a maximal entropy state were proven
to decay at least as fast as $1/t^{1/3}$\cite{centlim}.

For $N$ larger than $t/2$, the boundary conditions become important, as
one can see in the curves for $N=35$ and $N=51$, which start to show deviations
from the $N=101$ curve for times roughly $16$ and $22$ respectively.
This is no surprise, since the distance between the central spin and
the boundary is $N/2$, and $v_{sw}=1$ for $\Delta=0$.  We will see later
for $\Delta\neq 0$ that in general boundary effects become important
for $N/2=tv_{sw}$.
Interestingly, periodic boundary conditions offer no improvement, since
the $N=36$ periodic curve deviates at the same time as the $N=35$ open curve.

Another interesting effect is that once the boundary conditions become
important, the expectation value shows wild oscillations, which no longer
decrease in magnitude.  This may be a consequence of the fact that the
chain is integrable.  It would be interesting to see for a non-integrable
system
whether such oscillations occur or not; the simplest statistical assumption
for a non-integrable system is that the state at long times would be a
random pure state satisfying the conservation laws of total $S^z$ and total
energy, and thus the expectation value of $\langle S^z_i \rangle$ for
any $i$ would show only exponentially small fluctuations about
the average spin.

\begin{figure}
\centerline{
\includegraphics[scale=0.5]{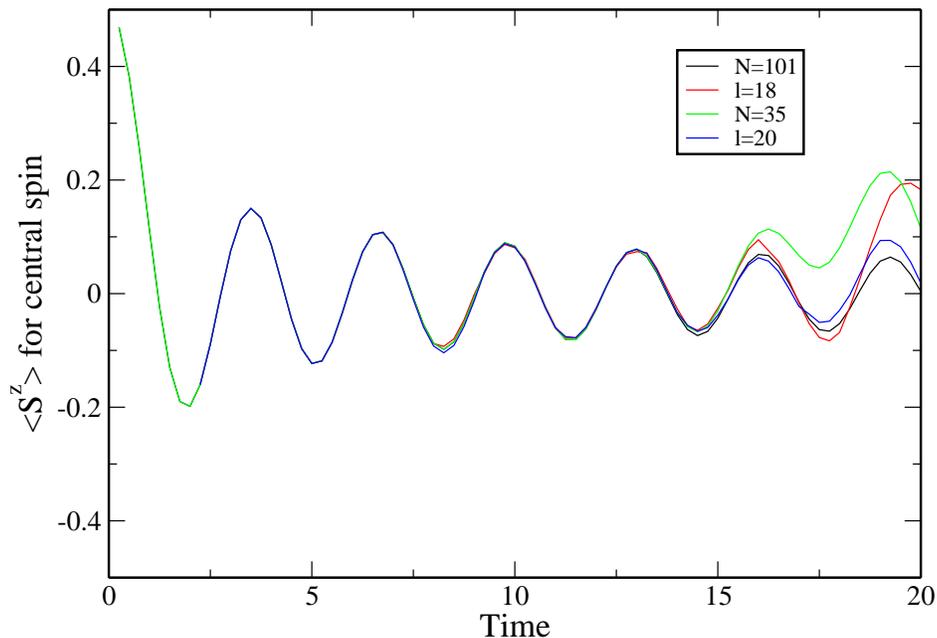}}
\caption{Time dependence of $\langle S^z \rangle$ for the central spin as
a function of time for $\Delta=0$.
Exact curves are $N=101$ (black) and $N=35$ (green).
Light-cone quantum circuit curves are
$l=18$ (red) and
$l=20$ (blue).
}
\label{d0}
\vspace{5mm}
\end{figure}

We now consider the application of the light-cone quantum
circuit method to this chain.  We considered $l=18$ and $20$, and
did $N_{it}=1000$ iterations of statistical sampling, as shown in Fig.~\ref{d0}.
The results for exact simulations with $N=35$ and $N=101$ are also shown
for comparison.  We see that the light-cone quantum circuit method with
$l=18$ is accurate over the same range of times as the exact simulation
with $N=35$, while the light-cone quantum circuit method with $l=20$
improves on this result.  By increasing $l$ from $18$ to $20$, we increase
the range of times by roughly $2$, while increasing $N$ by $2$ would increase
the range of times by only $1$.

\begin{figure}
\vspace{5mm}
\centerline{
\includegraphics[scale=0.5]{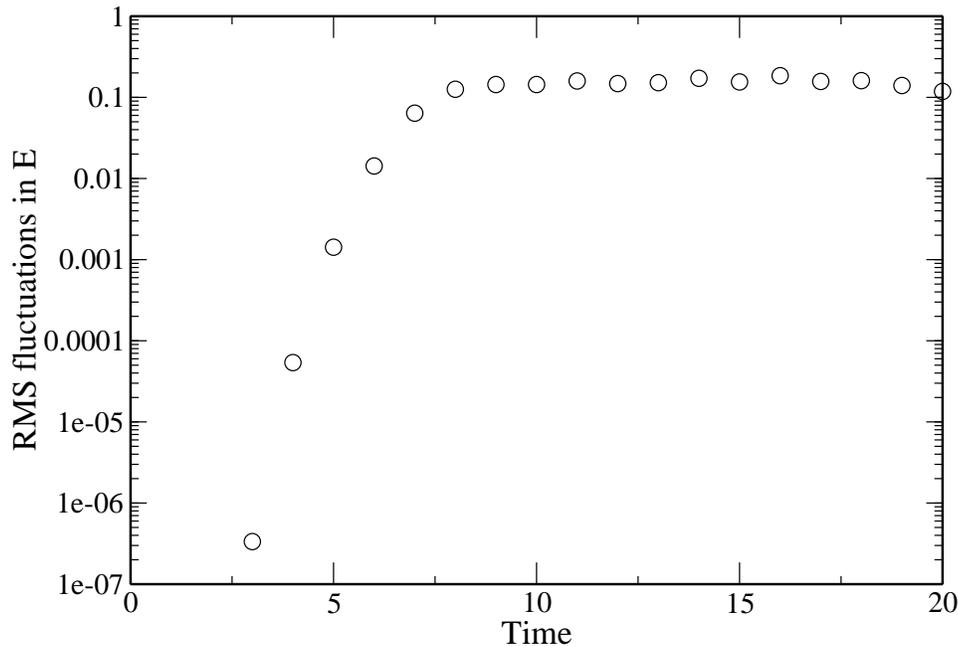}}
\caption{RMS fluctuations in $E(O,\alpha_L,\alpha_R)$ as described in
text as a function of time for $l=18$.}
\label{szav2}
\end{figure}

There are statistical fluctuations in the light-cone quantum
circuit results in Fig.~\ref{d0}, due to random
fluctuations in $E(O,\alpha_L,\alpha_R)$ for different choices of
$\alpha_L,\alpha_R$.  In Fig.~\ref{szav2} we plot the
rms fluctuation in $E(O,\alpha_L,\alpha_R)$ as a function of
time, sampling this expectation value
with the probability distribution $|A(\alpha_L)|^2 |A(\alpha_R)|^2$.
The results in Fig.~\ref{d0} are
an average over $N_{it}=1000$ samples, so the spread on each data point
in Fig.~\ref{d0} is equal to $1/\sqrt{1000}$ times the
fluctuation shown in Fig.~\ref{szav2}, or roughly $0.005$ in the worst
case.  A very interesting point is that for $t\leq 5$, the rms fluctuation
is negligible, while for times $t\geq 9$, the rms fluctuations
are up to their full value, with a sharp bend in the curve (plotted
on a log scale) around
$t=8$ or $t=9$.  This is a consequence of the maximum spin-wave
velocity: the influence of the regions on sites $-l,...,-l/2$ and $+l/2,...,+l$
takes a time of order $l/2v_{sw}$ to reach the central region.  There
are sill some fluctuations for $t\leq l/2v_{sw}=9$, but they decay rapidly
until they are negligible at very short time.

\begin{figure}
\centerline{
\includegraphics[scale=0.5]{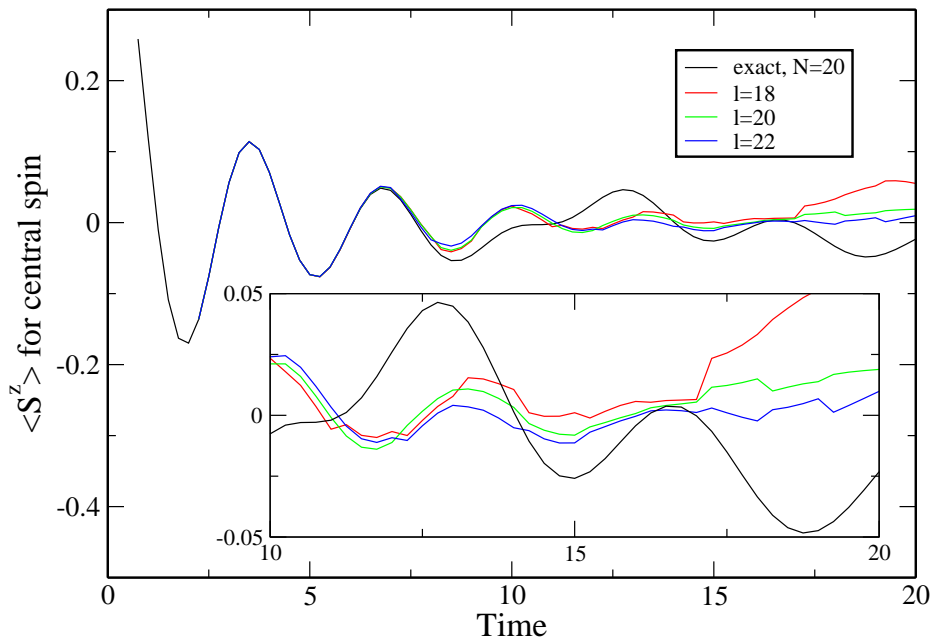}}
\caption{Time dependence of $\langle S^z \rangle$ for the central spin as
a function of time for $\Delta=0.5$.
Exact curve is $N=20$ (black).
Light-cone quantum circuit curves are
$l=18$ (red),
$l=20$ (blue), and
$l=22$ (green).
}
\label{d.5}
\vspace{5mm}
\end{figure}

\begin{figure}
\centerline{
\includegraphics[scale=0.5]{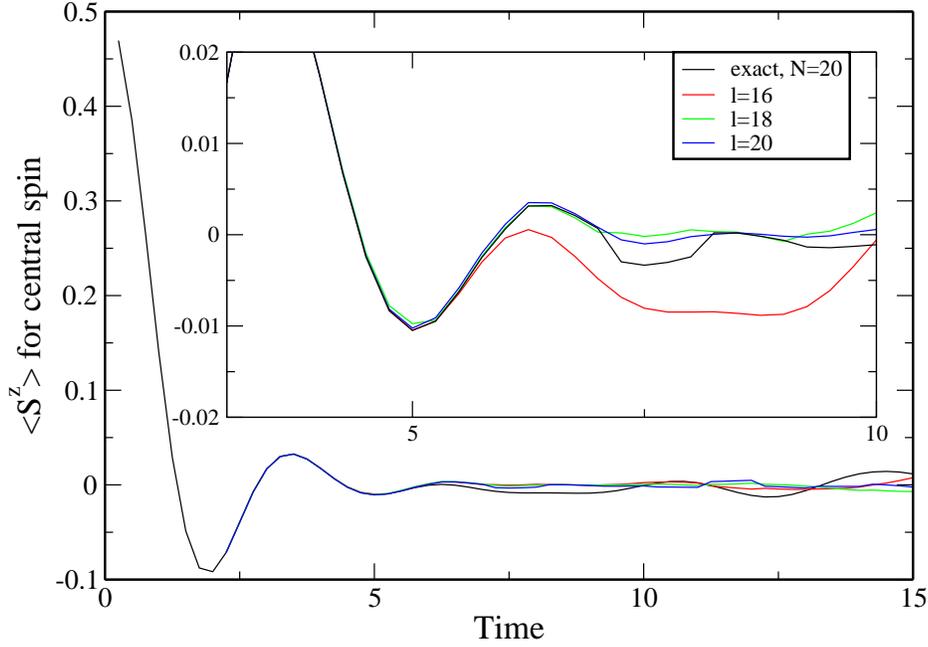}}
\caption{Time dependence of $\langle S^z \rangle$ for the central spin as
a function of time for $\Delta=1$.
Exact curve is $N=20$ (black).
Light-cone quantum circuit curves are
$l=16$ (red),
$l=18$ (green), and
$l=20$ (blue).
}
\label{d1}
\vspace{5mm}
\end{figure}

\begin{figure}
\centerline{
\includegraphics[scale=0.5]{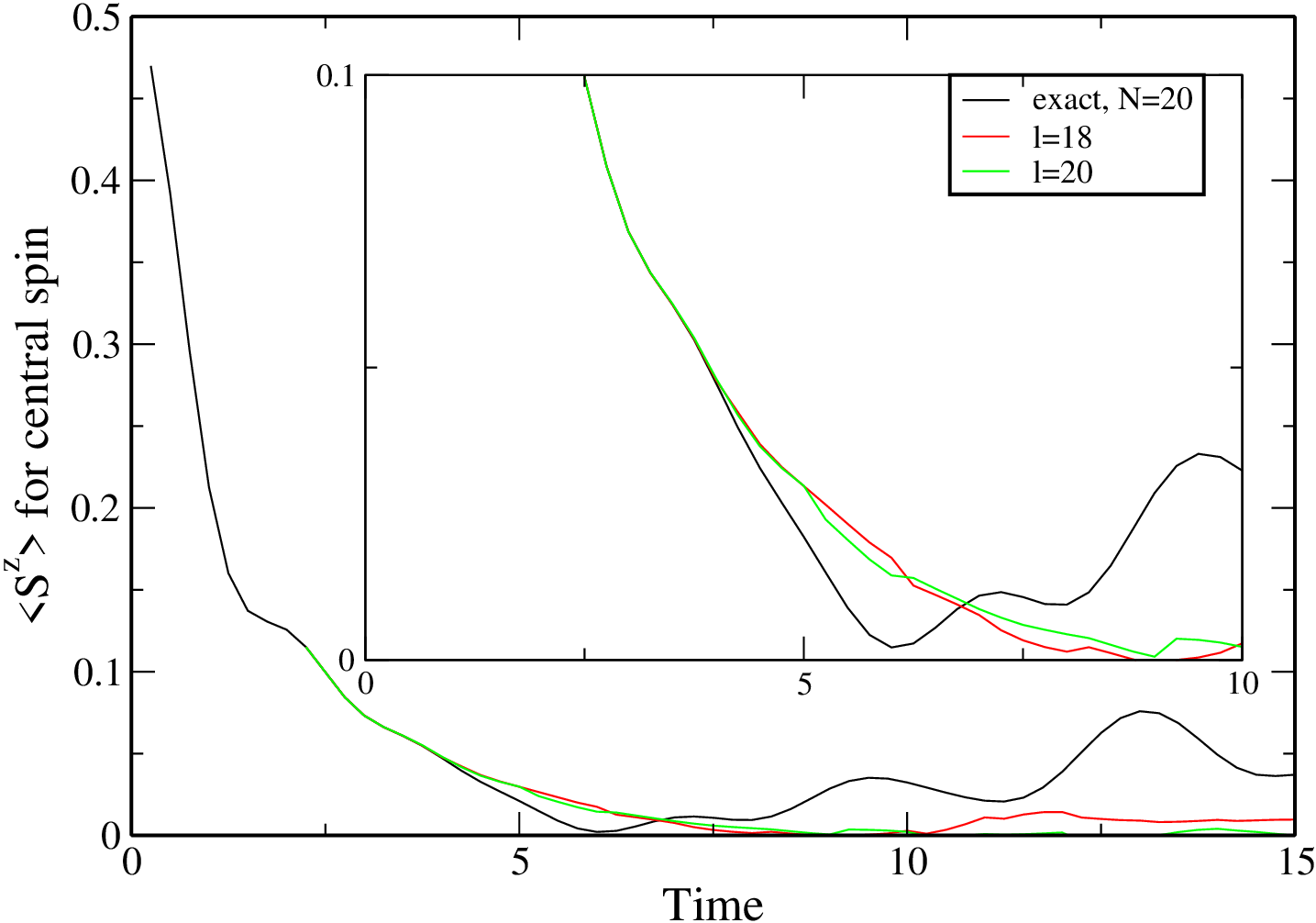}}
\caption{Time dependence of $\langle S^z \rangle$ for the central spin as
a function of time for $\Delta=2$.
Exact curve is $N=20$ (black).
Light-cone quantum circuit curves are
$l=18$ (red) and
$l=20$ (green).
}
\label{d2}
\vspace{5mm}
\end{figure}

We then applied the light-cone quantum circuit method to chains with
$\Delta=0.5,1,2$ as shown in Figs.~\ref{d.5},\ref{d1},\ref{d2}.
For comparison, we show an exact simulation of a chain with $N=20$.
For larger $\Delta=0.5,1$ we still see decaying oscillations, but the
decay is much more rapid than for $\Delta=0$.  The decay of the envelope
is, very roughly, $t^{-1.25}$ for $\Delta=0.5$.
For $\Delta=2$, no oscillations are seen.
The effects of statistical noise are much more noticeable, since the
magnitude of the spin is much less.  All simulations were done with
$N_{it}=1000$
statistical samples, except the simulation with $\Delta=0.5,l=22$ has only
$N_{it}=250$
samples, and the simulations with $\Delta=1,l=16$ and $\Delta=1,l=18$ had
$N_{it}=10000$ and $N_{it}=3000$ samples
respectively.  The spin-wave velocity is larger
for these chains, so the simulation breaks down at an earlier time than for
$\Delta=0$; we again see that the simulations work for
\be
t\lesssim N/(2v_{sw}),
\ee
or
\be
t\lesssim l/v_{sw},
\ee
in the exact and light-cone methods respectively.

\subsection{Entanglement Entropy}

From the predictions in \cite{cftrapid} and the numerical work in \cite{calab},
we can compare the difficulty of doing a similar calculation doing
time-dependent DMRG or time evolving bond decimation.  The entanglement
entropy is predicted to grow linearly in time.  In \cite{calab}, the prefactor
of the numerical growth was determined in a quench of an XXZ chain from
an Ising coupling with $\Delta=\Delta_0$ to an Ising coupling with strength
$\Delta_1$.  The prefactor depended on $\Delta_0$ and was largest in the
case of $\Delta_0=\infty$, the case we have considered here where the
initial condition is a Neel state.  There, in a quench to $\Delta_0=0$, the
entanglement entropy (with logs taken to base $2$) was observed to grow
a little faster than $0.6 v_{sw} t$.  This implies that the minimum size of the
bond dimension needed in a matrix product state is at least $2^{0.6 v_{sw} t}$,
which requires an effort going as $2^{1.8 v_{sw} t}$.  In contrast,
the numerical effort required to directly simulate a chain of length
$N$ scales as $2^N$.  To get accurate results for a time of order
$t$, requires $N=2 v_{sw} t$, or an effort $2^{2v_{sw} t}$.
Using the techniques
in the present paper, the effort can be reduced to 
of order $2^{v_{sw} t}$.

When comparing to \cite{calab}, note that there is a difference
in normalization of the Hamiltonian by a
a factor of 4, but since the results in \cite{calab} are expressed
in terms of the spin-wave velocity multiplied by time, the time axis
is the same for $\Delta=0$.  For $\Delta>0$, the time axis in \cite{calab}
differs from the time axis here,
since \cite{calab} multiplies the time by the relevant spin-wave velocity which
is greater than unity.
For $\Delta=1$, the spin-wave velocity using our normalization is equal
to $\pi/2>1$, and for $\Delta=0.5$ the spin-wave velocity is equal to
$3\sqrt{3}/4$, so the times in the present paper should be
multiplied by such a factor greater than unity when comparing to \cite{calab}.

In \cite{calab}, other initial conditions were considered, other than
just the Neel state.  These initial conditions were chosen to be
the ground state of various other XXZ Hamiltonians.  As the initial
$\Delta$ was reduced, the entropy growth was found to still be linear in time,
but with a smaller prefactor.  Such simulations could still be carried out
with our method as follows: first, use DMRG as done in \cite{calab} to find
the ground state on a long chain.  Then, suppose we are interested in
local observable on a region of length $l_0$.  To
find how these evolve
a time $t$ after a quench, locate some region of length $l_0+2 v t$ in the
chain.

We then ``observe" the state of the system outside this
region to statistically sample different pure states within the region,
and then evolve the pure states within the region.  This is
done as follows.  The DMRG ground state can be written in the form
\be
\label{3mps}
\Psi_0=\sum_{\alpha\beta} A_{\alpha\beta} |\Psi_L^{\alpha}\rangle\otimes
|\Psi_M^{\alpha\beta}\rangle\otimes |\Psi_R^{\beta}\rangle,
\ee
where $|\Psi_L^{\alpha}\rangle$ are a set of orthonormal states on
the chain to the left of the region, $|\Psi_R^{\beta}\rangle$ are a set
of orthonormal states to the right of the region, $\Psi_M^{\alpha\beta}$
are a set of states (normalized to unity but not necessarily orthogonal)
on the given region of length $l_0 + 2 v t$, and $A^{\alpha\beta}$
are a set of amplitudes.  We choose an $\alpha$ and a $\beta$ according to
the probability
\be
P=|A^{\alpha\beta}|^2,
\ee
and then evolve the state $\Psi_M^{\alpha\beta}$ using our present algorithm.
This corresponds to observing the matrix product state in the basis
$|\Psi_L^{\alpha}\rangle \otimes |\Psi_R^{\beta}\rangle$.
Repeating this statistical sampling
many times we obtain the desired quantities on the local region.
Note that on each
iteration we statistically sample $\alpha,\beta$ as well
as doing the sampling above.
The statistical sampling of the state outside the region may be
justified using Lieb-Robinson bounds as before.
Further, when a matrix product state is written in the canonical form,
the bond variables naturally have the orthonormality property used above to
do the statistical sampling, giving a state in the form (\ref{3mps}).

In some cases, if the entanglement entropy grows linearly in time but with
a sufficiently small prefactor, it may be worth using the light-cone
ideas above, but doing the initial
evolution for a time $t_i$ using matrix product methods instead of
quantum circuit methods, as follows.  Suppose the matrix product
methods require an
effort $t^2 2^{f t}$, for some number $f$, to simulate for a time $t$.
Then, to compute an observable at time $t_f$,
we simulate a subchain of length $2 v_{sw} t_f$ for a time $t_i$ using
matrix product methods.  We then statistically sample
states on a smaller subchain of length
$2 v_{sw} (t_f-t_i)$ and perform the simulation of that subchain
for time $t_f-t_i$ exactly.  The total
effort is then
\be
t^2 O(2^{f t_i}+2^{2v _{sw}(t_f-t_i})).
\ee
Choosing $t_i=t_f/(1+f)$ to minimize the computational cost, we find that
the cost scales as
\be
t^2 O(2^{f' t_f}),
\ee
with
\be
f'=\frac{1}{f^{-1}+(2v_{sw})^{-1})}.
\ee

\subsection{The Corner Transfer Quantum Circuit Method}

In this subsection we introduce the corner transfer quantum circuit method.
It is primarily of theoretical, rather than practical interest.  For
calculation of local quantities (such as the expectation value of the
spin on a single site) the light-cone method above is less work.  However,
the corner transfer method does give an approximation to the full wavefunction,
and may be less work than variational
matrix product methods in cases where the entanglement
entropy grows rapidly.

We now define the quantum circuit that approximates $\exp(-i H t)$.
We define a length scale $l'<<v_{LR} t$:
the error in our quantum circuit approximation
to $\exp(-i H t/2)$ will be exponentially small in $l'$, while the
maximal velocity of information propagation for our quantum circuit will
be 
\be
\label{vmax}
v_{max}=
v_{LR}+{\cal O}(l'/v_{LR} t).
\ee

\begin{figure}
\centerline{
\includegraphics[scale=0.5]{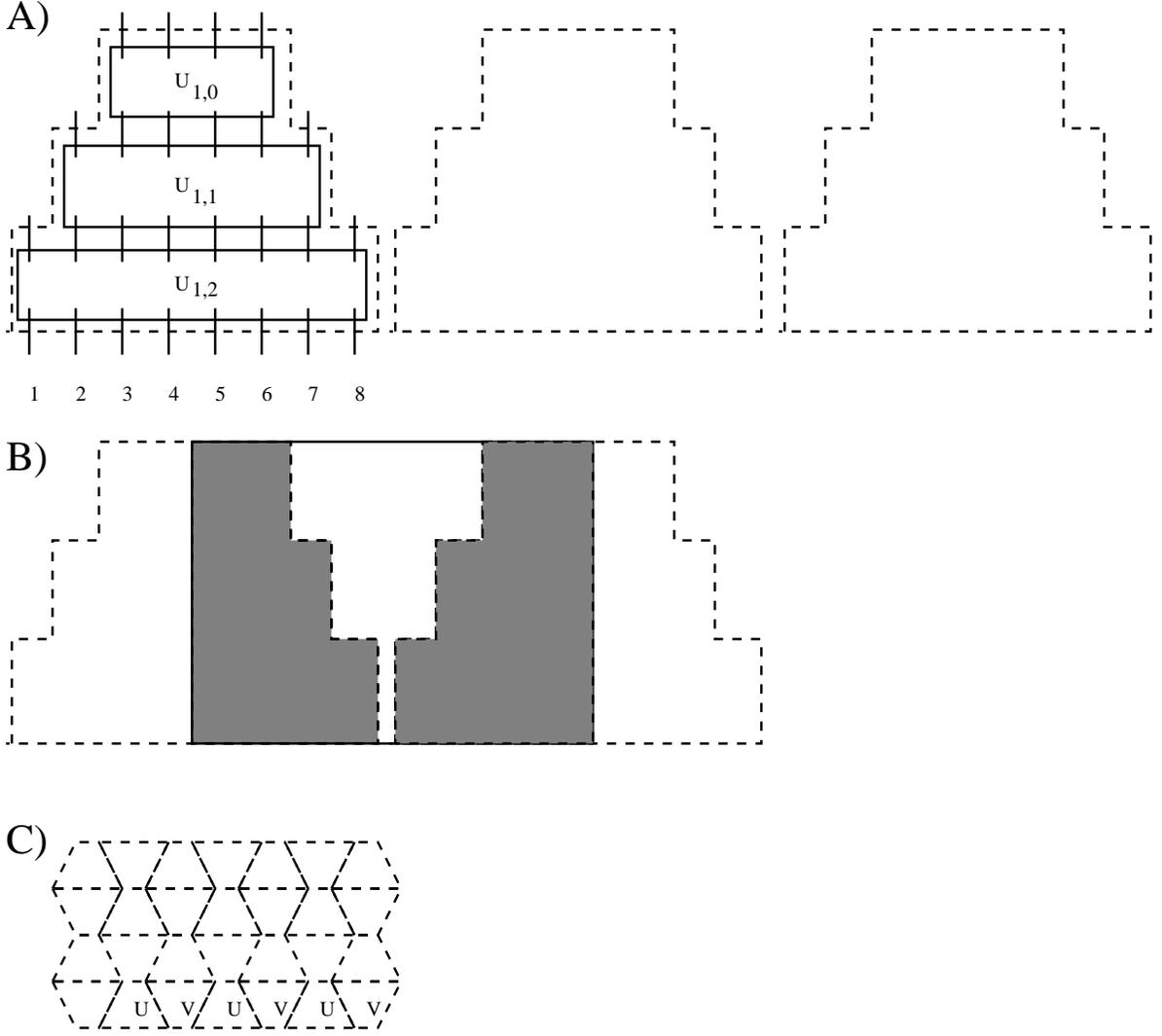}}
\caption{(a) The dashed lines show the
region of space-time used in defining
operators $U_k$, with space on the horizontal axis and time on
the vertical axis.  Only three such regions are shown, but
the pattern repeats over the entire system.
We also show this as a quantum circuit writing
$U_1=U_{1,0},U_{1,1},U_{1,2}$ where each operator $U_{1,n}$ computes
the exponential of the Hamiltonian in a given time slice.  Construction
is shown for $l'=2,n=3$.
(b) Action of operators $V_k$ as discussed in text.
(c) Iterating many rounds of the corner transfer quantum circuit.
On the bottom row we label $U$ and $V$; on the row above, a $\tilde U$
sits above each $V$ and a $\tilde V$ above each $U$.}
\label{ctqc}
\vspace{5mm}
\end{figure}

We construct the quantum circuit in two steps, first
presenting a quantum circuit that approximates $\exp(-i H t/2)$.
We introduce operators $U_k$ that describe the time evolution under
a time dependent Hamiltonian: each $U_k$ contains at a time $t$ only
the interaction terms which are contained within one of the triangles with
a flattened top and jagged sides surrounded by a dashed line shown
in
in Fig.~\ref{ctqc}(a).  That is, we break the time $t/2$
into $n_0=\lceil v_{LR} (t/2)\rceil$ subintervals of time at most
$1/v_{LR}$.  We place the center of the triangles on sites $kl$, for
$k$ integer, where
\be
l=2 l'+v_{LR} t.
\ee
Then we define $U_{k,n}$ for $0\leq n\leq n_0-1$ by
\be
U_{k,n}=
\exp[-i \sum_{i=kl-l'/2-n}^{i\leq kl+l'/2+n} h_i (t/2n_0)],
\ee
and define
\be
\label{before}
U_k=U_{k,0} U_{k,1} U_{k,2} ... U_{k,n_0-1}.
\ee

We then define the operators $V_k$ as follows.  We define
\begin{eqnarray}
U_{k,n}^L=
\exp[-i \sum_{i=kl-l'/2-n}^{kl} h_i (t/2n_0)], \\ \nonumber
U_{k,n}^R=
\exp[-i \sum_{i=kl}^{kl+l'/2+n} h_i (t/2n_0)].
\end{eqnarray}
That is, $U_{k,n}^{L,R}$ represent evolution under the left or right
half of the triangles in in Fig.~(\ref{ctqc}(a).  Then, we define
\be
\label{analog}
V_k=
\exp(-i \sum_{i=(k+1/2)l-l/2+1}^{(k+1)l+l/2-1} h_i t/2)
\Bigl( U^R_{k,0} U^R_{k,1} ... U^R_{k,n_0-1}\Bigr)^{\dagger}
\Bigl( U^L_{k+1,0} U^L_{k+1,1} ... U^L_{k+1,n_0-1}\Bigr)^{\dagger}.
\ee
That is, each $V_k$ ``undoes" the evolution in the shaded triangles as shown
in Fig.~\ref{ctqc}(b), and then performs the full evolution in the rectangle
bordered by the solid line.

We now approximate
\be
\exp(-i H t/2)\approx \prod_k V_k \prod_k U_k.
\ee
Using Lieb-Robinson bounds, one can show that the error in this approximation
is
\be
\Vert
\exp(-i H t/2)-\prod_k V_k \prod_k U_k
\Vert
\leq {\cal O}(t \sum_i \Vert h_i \Vert \exp(-{\cal O}(l')).
\ee

In the second step of construction the corner transfer quantum circuit,
we define another approximation to $\exp(-i Ht /2)$.  
We set
\be
\tilde U_{k,n}=
\exp[-i \sum_{i=(k+1/2)l-l'/2-n}^{i\leq (k+1/2)l+l'/2+n} h_i (t/2n_0)],
\ee
and define
\be
\tilde U_k=\tilde U_{k,0} \tilde U_{k,1} \tilde U_{k,2} ... \tilde U_{k,n_0-1}.
\ee
Here, the
centers of the upward facing triangles in
Fig.~\ref{ctqc}(a) are shifted by $l/2$, compared to (\ref{before}).
We then define $\tilde V_k$ in analogy to Eq.~(\ref{analog}) with the
centers again shifted by $l/2$ and approximate
\be
\exp(-i H t/2)\approx \prod_k \tilde V_k \prod_k \tilde U_k.
\ee

In
Fig.~\ref{ctqc}(c) we show multiple rounds of
the quantum circuit.  Except for the row of triangles
on the top and bottom boundaries, the quantum circuit looks like
diamond-shaped patches in space time.  They are rotated 45-degrees from
the patches in \cite{qcirc}, justifying the name ``corner transfer";
the 45-degree rotation is the key improvement over \cite{qcirc}
leading to the bound on information propagation (\ref{vmax}).

\section{Quantum Belief Propagation}

In this section we apply quantum belief propagation to disordered
systems.  We begin with a brief review of quantum belief propagation,
focusing on the computational effort required, and then discuss
modifications for disordered systems.  We applied this procedure to
two different disordered systems: one a chain with no frustration where
Monte Carlo is available for comparison\cite{FAF}, and the other a frustrated
spin system with disorder\cite{FrusDis}.

Quantum belief propagation\cite{qbp} is a method for constructing
a matrix product density operator for a thermal state of a quantum
system on a line or other loopless lattice.  The algorithm depends on
a parameter $l_0$ and the approximation to the thermal state has the form
in one dimension of:
\be
\rho=\Bigl( O_{N-l_0+1}^{\dagger} ... O_2^{\dagger} O_1^{\dagger} \Bigr) \rho_0 
\Bigl(O_1 O_2 ... O_{N-l_0+1}\Bigr),
\ee
where the operators $O_i$ act on sites $i,...,i+l_{0}-1$ and the density
matrix $\rho_0$ has the form $\rho_0=\rho_0^{1...l_0} \otimes \openone
\otimes ... \otimes \openone$, with $\rho_0^{1...l_0}$ being a density
matrix on sites $1...l_0$.  The implementation of the algorithm
depends on tracing out sites, a process analogous to the observations
discussed above.  Suppose we wish to compute 
the partition function.
If all of the $O_i$ are the same, as they would be
in a system without disorder, then we can consider the completely positive
map
\be
\label{CPmap}
\rho \rightarrow tr_1(O_i^{\dagger}\rho O_i) \otimes \openone),
\ee
where $tr_1(...)$ denotes a trace over the first site, and $\rho$ is a density
operator on an interval of length $l_0$.  We start with this map at
$\rho=\rho_0$, and iterate it until we reach the end of the chain.
We then compute the trace of the final $\rho$ and this is the
of the partition function.  A similar procedure can be applied to
compute expectation values.

The cost of the procedure scales exponentially in $l_0$, as it requires
diagonalizing matrices of dimension $2^{l_0}$\cite{state}.  The procedure
is effective down to inverse temperatures
\be
\beta\sim l_0/v_{LR}.
\ee
The physical intuition is that the algorithm keeps quantum effects only
up to a length scaling $l_0$.  However, as we will see, the algorithm
is capable of keeping track of classical correlations on much longer length
scales; this should be no surprise since for classical Hamiltonians which are
the sum of commuting operators,
such as the Ising Hamiltonian $H=\sum_i S^z_i S^z_{i+1}$, quantum belief
propagation exactly reproduces classical transfer matrix methods which
can exhibit correlation lengths exponentially large in $\beta$.

For a disordered system, the operators $O_i$ may vary as a function of $i$,
but they depend only on the bonds on sites $i...i+l_0-1$.  Below, we consider
a system in which the bonds can assume only discrete values; in the first
system there are $2\cdot 2^{(l_0-1)/2}$ different possible choices for the set
of bonds
on sites $i...i+l_0-1$ while in the second there are $2^{l_0-1}$
possible choices.  We then use the following algorithm: first, we pre-compute
the operator $O$ for each possible choice of bonds.  We then randomly generate
a configuration of bonds, and iterate the map (\ref{CPmap}) above,
choosing the appropriate $O$ at each step.  This requires an effort which scales
linearly in system size.

Interestingly, the algorithm seems to work better at low temperatures on
disordered systems than on ordered systems.  The reason is probably
the following: when deriving the algorithm in \cite{qbp}, we used Lieb-Robinson
bounds with velocity $v_{LR}$.  However, just as in the non-equilibrium
case above where the actual velocity $v_{sw}$ is less than $v_{LR}$, in these
thermal systems the actual velocity may again be less than $v_{LR}$.  For
systems with disorder, localization effects may further reduce the velocity,
and even change the ballistic spreading of the wavepacket to a slower growth.
Interestingly, this phenomenon is known by different names in condensed
matter, where it is called many-body localization\cite{alt}, and quantum
information theory, where it is called a Lieb-Robinson bound for a disordered
system\cite{LRdis}.

\subsection{Results on Disordered Systems}

\begin{figure}
\centerline{
\includegraphics[scale=0.5]{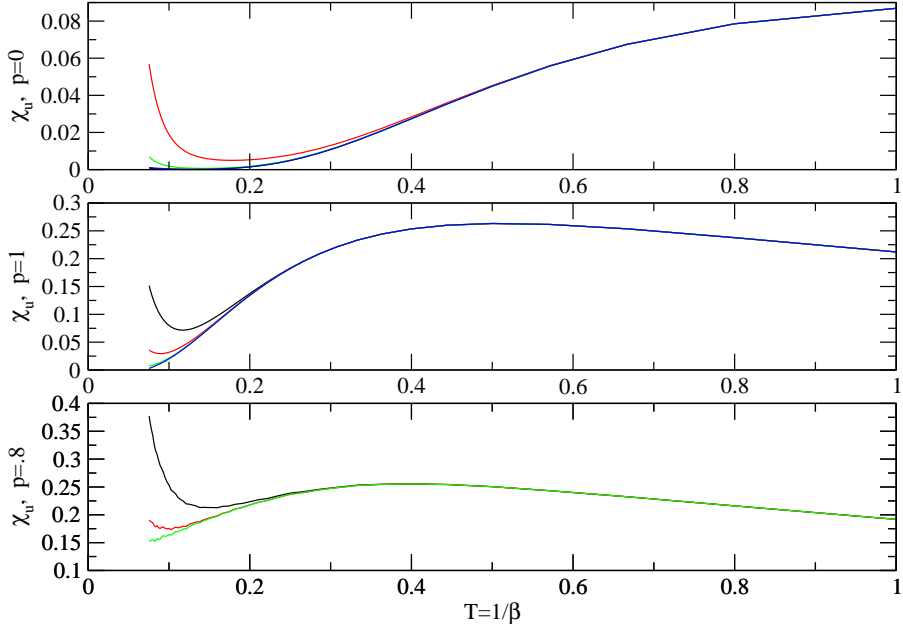}}
\caption{Uniform susceptibility as a function of temperature
for different $p,l_0$.  
Top: $l_0=5$ (black), $l_0=7$ (red), $l_0=9$ (green).
Middle: $l_0=5$ (black), $l_0=7$ (red), $l_0=9$ (green), $l_0=11$ (blue).
Bottom: $l_0=5$ (black), $l_0=7$ (red), $l_0=9$ (green), $l_0=11$ (blue).
}
\label{faffig}
\vspace{5mm}
\end{figure}

\begin{figure}
\centerline{
\includegraphics[scale=0.5]{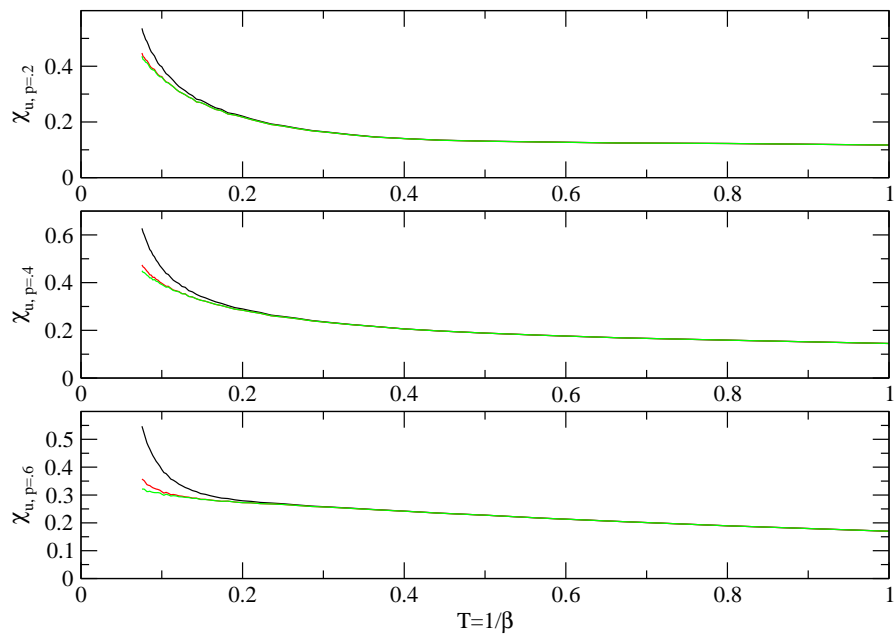}}
\caption{Uniform susceptibility as a function of temperature
for $p=0.2,0.4,0.6$ from top to bottom and for
$l_0=5$ (black), $l_0=7$ (red), $l_0=9$ (green).
}
\vspace{5mm}
\label{faffig2}
\end{figure}

The first disordered system we considered has the Hamiltonian
\be
H=\sum_{i=1}^{N/2} J \vec S_{2i-1} \cdot \vec S_{2i} + \sum_{i=1}^{N/2}
J_i S_{2i}\cdot S_{2i+1},
\ee
where each spin is spin-$1/2$, $J>0$ and $J_i=J_F<0$ with probability
$p$ and $J_i=J_A>0$ with probability $1-p$.  This model has
both ferromagnetic and anti-ferromagnetic couplings, and was proposed
to model the compound $(CH_3)_2CHNH_3Cu(Cl_xBr_{1-x})_3$, where
the probability $p=x^2$\cite{c1,c2,c3,c4,c5}.  The case considered is
$J=1$ and $|J_i|=2$.  We performed simulations on this chain with
$l_0=5,7,9$ for $p=0.2,0.4,0.6,0.8$ and for $l_0=5,7,9,11$ for
$p=0,1$.  For the random cases ($p=0.2,0.4,0.6,0.8$) we considered chains
of $100000$ sites and computed the uniform susceptibility by the change
in partition function in response to a weak applied field.  In this way,
the susceptibility was self-averaging.  For the
pure cases, we considered shorter chains, and computed the applied
susceptibility by measuring partition functions as in \cite{qbp}.
The results are shown in Figs.~\ref{faffig},\ref{faffig2}.
For small $l_0$, qualitatively
wrong results are seen at low temperature, with the susceptibility diverging
at low enough temperature.  However, as $l_0$ is increased, the accuracy
extends to lower temperature.  The difference between the curves for
$l_0=7$ and $l_9$=9 is small for $T$ greater than roughly $1/7$.
In this region also, we agree well with Monte Carlo data.

Data was taken over several different $\beta$, with a step of $0.25$ in
$\beta$.  As in \cite{qbp}, we set the perturbation $A$ (following notation
of \cite{qbp}) to equal $(1/2)(h_{l_0-2,l_0-1}+h_{l_0-1,l_0})$ rather
than $A=h_{l_0-1,l_0}$.

The second model we considered is \cite{FrusDis} a model with frustration
and disorder, where quantum Monte Carlo results are not available.
The study in \cite{FrusDis} was motivated by experimental studies
on $CuGeO_3$\cite{exper} and $Cu_6Ge_6O_{18}-xH_20$\cite{exper2},
where second neighbor interactions may be important.
We were interested in
a case where second neighbor interactions would be very important, so we
considered the Hamiltonian
\be
H=\sum_{i=1}^{N-1} J_i \vec S_i \cdot \vec S_{i+1} +
\sum_{i=1}^{N-2} K_i \vec S_i \cdot S_{i+2},
\ee
and in the pure case we considered $J_i=1,K_1=1/2$.  This is a Majumdar-Ghosh
chain with a dimerized ground state.  In the disordered case we chose
$J_i=0.9$ or $J_i=1.1$, with probability $1/2$ of either choice
(similar results were found for choosing $J=0.75$ or $J=1.25$).  We choose
\be
K_i=(1/2) J_i J_{i+1},
\ee
which correlates the second neighbor interaction with the nearest neighbor
interaction as described in \cite{FAF}.

\begin{figure}
\centerline{
\includegraphics[scale=0.5]{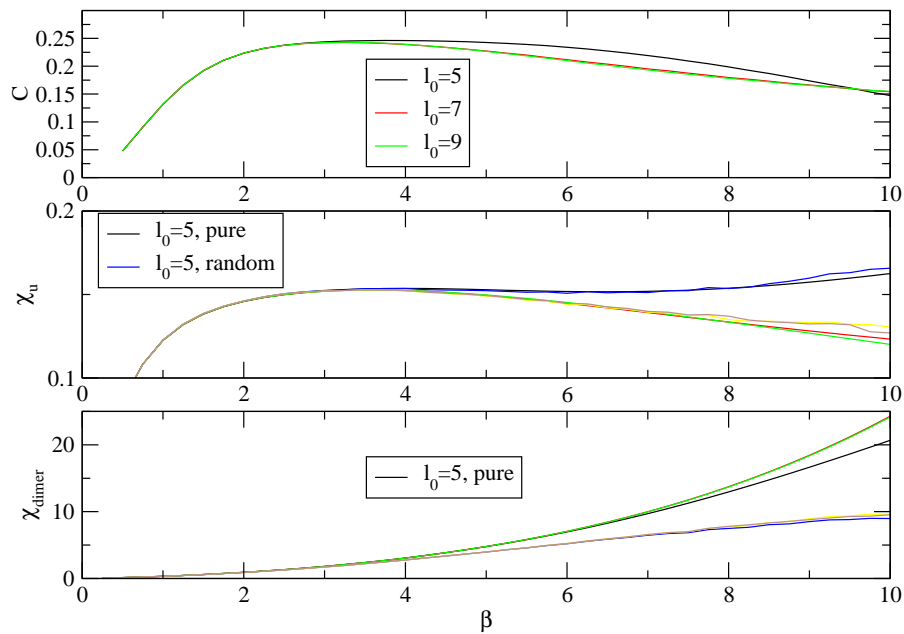}}
\caption{Top: specific heat for the pure system.
Middle: uniform susceptibility for
the pure system ($l_0=5,7,9$ are black, red, green respectively)
and the disordered system ($l_0=5,7,9$ are blue, yellow, brown respectively).
Bottom: dimer susceptibility for
the pure system ($l_0=5,7,9$ are black, red, green respectively)
and the disordered system ($l_0=5,7,9$ are blue, yellow, brown respectively).
}
\label{FDfig}
\vspace{5mm}
\end{figure}

We studied $l_0=5,7,9$ with chains of length $19999$.  It is necessary
to take such long chains in the pure case to avoid boundary condition
effects because at low temperatures in the pure system there is an exponentially
increasing correlation length for dimer-dimer correlations.
In Fig.~\ref{FDfig}) we first show the results of the specific heat
as a function of $\beta$
for the pure case,
computed from the second derivative
of the partition function (probably a very slightly
more accurate method is to take the
first derivative of the energy as in \cite{qbp}).  A strong difference is
seen between $l_0=5$ and $l_0=7$ above $\beta\sim 3.25$, but the
$l_0=7$ and $l_0=9$ curves are almost identical.  This indicates that
by going to $l_0=9$ we have succeeded in converging the
specific heat in $l_0$ for $\beta\leq 10$.

The uniform susceptibility shows a similar effect.  The pure curves
show a large difference between $l_0=5$ and $l_0=7,9$, but only slight
difference can be seen between $l_0=7,9$ and only above $\beta=8$.  Again,
the results seem to be converged in $l_0$ by going to $l_0=9$ in the range
of temperatures we consider.
The disordered curves show again that $l_0=5$ is too small, but for $l_0=7,9$
little difference is seen (except for some small random fluctuations) up
to $\beta=10$.  A very slightly higher susceptibility is seen in the
random case compared to the pure case.
Finally, we consider the dimer susceptibility, defined by
\be
\chi_{dimer}=\beta \langle \Bigl( \sum_{i=1}^{N/2-1} \vec S_{2i} \cdot
\vec S_{2i+1}-\vec S_{2i+1} \cdot \vec S_{2i+2} \Bigr)^2 \rangle.
\ee
This shows a large difference between the pure and disordered cases.
The pure cases again show agreement between $l_0=7,9$ and show an an increase
that gives $\chi_{dimer}/\beta$ growing exponentially in $\beta$.
The disordered cases show $\chi_{dimer}/\beta$ saturating as a function
of $\beta$.

The saturation of the dimer-dimer correlation function in the disordered
case is no surprise.  The disorder locally breaks the $Z_2$ symmetry between
different ordering patterns of the dimers, and is relevant for this one
dimensional system.  The fact that the uniform susceptibility shows
only very slight difference between pure and disordered systems
is more surprising; asymptotically, the uniform susceptibility should
decay exponentially in $\beta$ in the pure system and should increase
as $\beta/\log^2(\beta)$ in the random singlet phase\cite{FrusDis}.

\section{Discussion}

The main result in this paper is the light-cone quantum circuit method.
We have tested this method numerically on a free system, with $\Delta=0$,
and on interacting systems with $\Delta\neq 0$.  We have found decaying
oscillations in the expectation value of the spin.  In future, this
technique will be useful for studying non-integrable systems, to
see if they relax to a thermal state\cite{sb}.

We can study two-dimensional systems by considering them as wide one-dimensional
systems; this allows us to double the number of spins, but only leads to
a factor of $\sqrt{2}$ increase in the time compared to direct simulation.
Other similar quantum circuit methods may be more effective in two
dimensions.

The results using the light-cone quantum circuit method are indeed
comparable to
those one would find by exactly solving a system of twice the size,
at least for $\Delta=0$ where we can compute exactly.
Thus, we $l=18$, we find results comparable to a system of $N=35$ or
$N=37$ sites.
The computational cost to exactly evolve a given system is roughly
comparable to that
required to do exact diagonalization using Lanczos
methods on that system: Lanczos methods and exact evolution both
require sparse matrix-vector multiplication, but the number of multiplications
needed to reach convergence for the time evolution may be larger than
that needed to reach convergence for ground state properties.
Thus, we expect that sizes around $35$ sites,
especially given the low symmetry of the present system, are around
the
upper limit for exact methods now, while we carried the light-cone quantum
circuit method up to $l=22$.  Further, the asymptotic
analysis of time requirements applies also to memory requirements:
the memory requirements of an exact solution on $N$ sites scale
as $N2^N$, while
while the memory requirement
of the light-cone quantum circuit method scale as $l 2^l$, so,
regardless of what $N$ can be obtained using an exact solution, it should
be possible to obtain the same $l$, up to a difference of a couple sites,
in the light-cone quantum circuit method.  The main additional cost in the
light-cone quantum circuit method is the need to run many times to obtain
statistical samples, but this is a problem which can be parallelized.

{\it Acknowledgments---} I thank R. Melko for useful comments on implementing
sparse matrix-vector multiplication.  I thank the KITP for hospitality
while this research was conducted.
This research was supported in part by the National Science  
Foundation under Grant No. PHY05-51164.
This work supported by U. S. DOE Contract No. DE-AC52-06NA25396.


\begin{thebibliography}{99}
\bibitem{2n2why} To understand why $2^{N/2}$ states suffice, consider
how much entanglement is possible across each bond.  If we divide the chain
of $N$ sites into a subchain of $n$ sites and another subchain of $N-n$
sites, with $n\leq N-n$, then there are only $2^n$ states available on the
subchain and so it is possible to perform a Schmidt decomposition and
find a matrix product state representation with $2^n$ states across that bond.
The worst case is midway across the chain, when $n=N/2$.  If we wish to
write matrix product operators instead of states, then these estimates are
increased to $4^{N/2}$.
\bibitem{spec} The Krylov space methods may lead to some improvement
in the time required compared to the series methods.
However, since we are considering a situation in which the entropy
increases rapidly, and hence the state $\exp(i H_L t)\Psi_L$ has very little
overlap with $\Psi_L$, it seems likely that it would be necessary to take
a fairly large $k$ to obtain accurate results for large $t$.  Therefore,
we are not sure how much advantage would actually be obtained by
using Krylov methods here.
\bibitem{state} The quantum belief propagation method, as currently
implemented, requires manipulating operators, rather than pure states, which
increases the computational effort.
\end{thebibliography}
\end{document}